\documentclass[conference]{IEEEtran}
\usepackage{cite}
\usepackage{amsmath,amssymb,amsfonts}
\usepackage{algorithmic}
\usepackage{graphicx}
\usepackage{textcomp}
\usepackage{makecell}
\usepackage{booktabs}
\usepackage{xcolor}
\begin{document}

\title{Beamforming with Random Projections: Upper and Lower Bounds}

\author{
\IEEEauthorblockN{Manan Mittal\IEEEauthorrefmark{2}, Ryan M. Corey\IEEEauthorrefmark{3}, Andrew C. Singer\IEEEauthorrefmark{2}}

\IEEEauthorblockA{\IEEEauthorrefmark{2}
Stony Brook University, Stony Brook, New York}
\IEEEauthorblockA{\IEEEauthorrefmark{3}
University of Illinois Chicago and Discovery Partners Institute, Chicago, Illinois}
}
\newtheorem{theorem}{Theorem}[section]
\newtheorem{corollary}{Corollary}[theorem]
\newtheorem{lemma}[theorem]{Lemma}
\maketitle

\begin{abstract}  
Beamformers often trade off white noise gain against the ability to suppress interferers. With distributed microphone arrays, this trade-off becomes crucial as different arrays capture vastly different magnitude and phase differences for each source. We propose the use of multiple random projections as a first-stage preprocessing scheme in a data-driven approach to dimensionality reduction and beamforming. We show that a mixture beamformer derived from the use of multiple such random projections can effectively outperform the minimum variance distortionless response (MVDR) beamformer in terms of signal-to-noise ratio (SNR) and signal-to-interferer-and-noise ratio (SINR) gain. Moreover, our method introduces computational complexity as a trade-off in the design of adaptive beamformers, alongside noise gain and interferer suppression. This added degree of freedom allows the algorithm to better exploit the inherent structure of the received signal and achieve better real-time performance while requiring fewer computations. Finally, we derive upper and lower bounds for the output power of the compressed beamformer when compared to the full complexity MVDR beamformer. 
\end{abstract}


\section{Introduction}

Acoustic sensor networks capture spatially varied information, with nodes consisting of acoustic sensors, such as microphone arrays, connected over a communication network. This configuration enables a variety of applications, including speech enhancement and source localization \cite{6101302}. However, in real-world environments, the desired signal is often contaminated by interfering signals and background noise. Beamforming algorithms leverage the spatial distribution of microphone elements to separate and enhance the desired signal from the captured mixture of signals \cite{7805139, 10248188, 9022475, corey22_interspeech}. For a real-time system, the use of all microphones in a beamforming system, which is referred to as \textit{sensorspace} processing, may not exploit the inherent sparsity of the received signals. This can occur when the number of sources is much smaller than the number of microphones. In such cases, the received signal is sparse with respect to the set of acoustic transfer functions in the environment considered. When received signals are sparse in an appropriate sense, previous work has demonstrated the effectiveness of using compressive sensing to perform signal recovery \cite{candes_romberg, candes_tao, candes_decoding, donoho, angeliki, edelmann}. Under the assumption that the received signal is sparse in some fixed basis, compressive sensing tries to recover the signal using a finite number of observations, typically much smaller than the dimension of the signal \cite{candes_romberg, davenport_duarte}. Measurements are obtained using a sensing matrix.  The sensing matrix, many of them random \cite{candes_tao, baraniuk}, must satisfy the Restricted Isometry Property (RIP). Candes and Tao \cite{candes_tao} show that solutions satisfying their proposed $\ell_1$ minimization, have bounded error in the $\ell_2$-sense, as well. However, for a general room environment, it can be difficult to have access to a dictionary of known acoustic transfer functions or construct an appropriate environment informed sensing matrix. Additionally, for real-time applications iterative procedures required for estimates may incur a high latency.

\begin{figure}
	\centerline{\includegraphics[width=8cm]{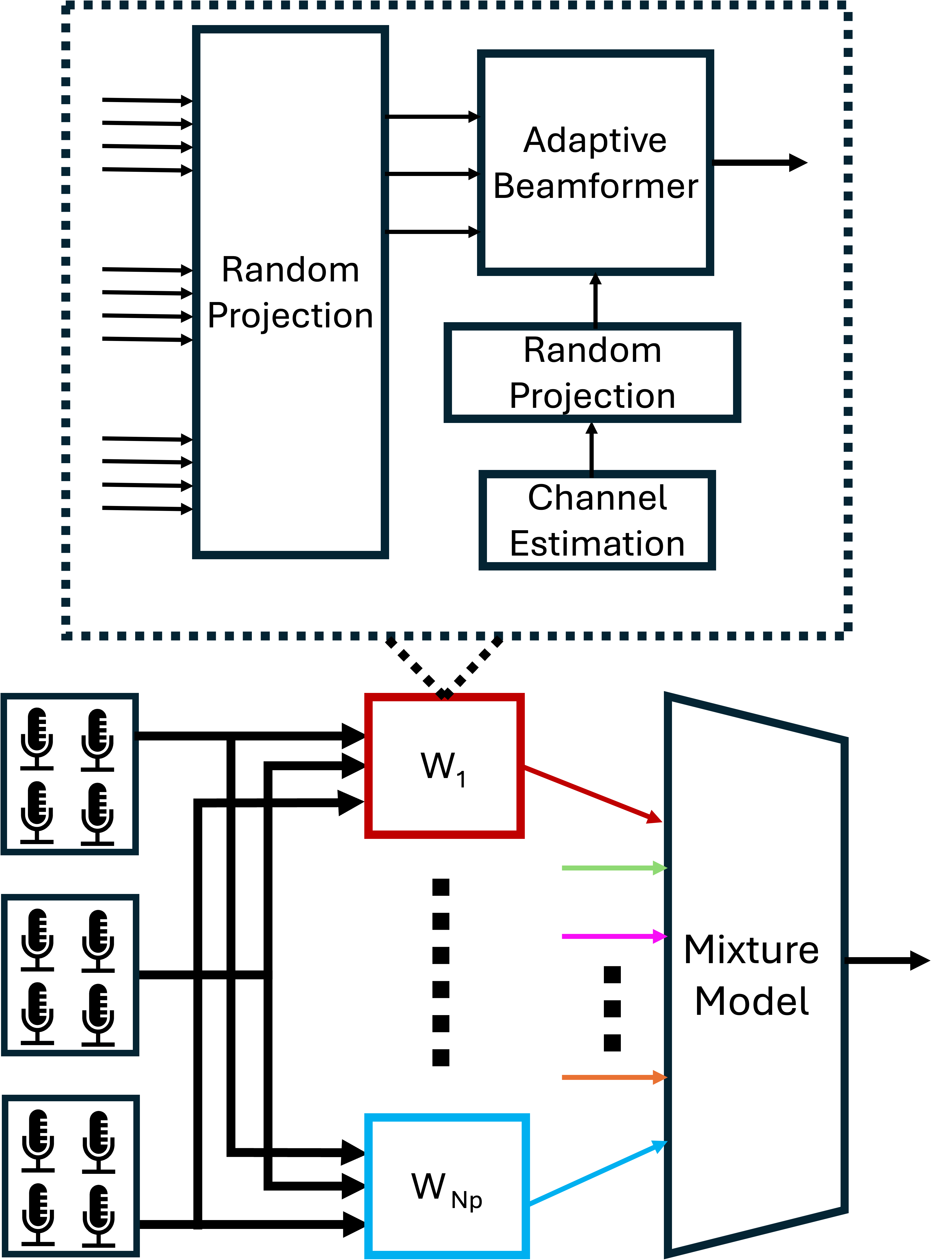}}
	\caption{A system diagram of the proposed method. The signals from each microphone array are processed using several compressed beamformers. Each beamformer has a different random projection and computes a projected channel estimate to use in an adaptive beamformer. Those outputs are in turn combined using a mixture model that is generally dependent on the type of signal being estimated.}
	\label{fig:compressed_beamformer}
\end{figure}

A common approach to reduced dimension beamforming is to preprocess the signals from the full array with a smaller number (than the array size) of conventional beamformers whose outputs are then processed using traditional signal-dependent beamforming methods such as the minimum variance distortionless response (MVDR) beamformer \cite{iwaenc_manan, stojanovic_jasa, stojanovic_oceans, friedlander, feldman}.  In \cite{davenport_boufounos} common signal processing tasks are solved using compressed sensing techniques (random projections), but using $\ell_2$-minimization instead of the iterative $\ell_1$-minimization algorithms. This is closely related to the Johnson-Lindenstrauss (JL) lemma, which asserts the existence of bi-Lipschitz mappings between a high-dimensional signal space and a lower-dimensional embedding space. In this work, we employ JL compatible matrix constructions to beamforming. In particular, we propose a system that can effectively reduce the dimensionality of the problem and compensate for any distortion induced by projection into a lower-dimensional space. The distinction is that our method parallelly employs multiple projections and mixes the resultant beamformers in a data-driven sense. The result demonstrates that more realizations of smaller projections may achieve better performance than one realization of a larger projection. Figure \ref{fig:compressed_beamformer} depicts the proposed system with an illustration of the internal structure of each compressed beamformer. 

Our method allows a system designer to take advantage of both forms of sparsity present in the acoustic environment. The compressed beamformer allows us to exploit the sparsity of the talkers distributed in the room (or equivalently the sparsity of the received signal), while the mixture model allows us to exploit the time-frequency sparsity of speech. We demonstrate through our experiments that our method achieves higher SNR and SINR gain than the conventional MVDR beamformer in a simulated room environment, where all signals are assumed to be zero mean.  

\section{Signal Model}
Consider a group of $N_s > 1$ talkers and $N_m$ microphones. We assume that the $N_m$ microphones are distributed across a room with arbitrary geometry. As such, we assume that they can share signals freely, where a centralized processing unit performs the required beamforming. We consider the short time Fourier transform (STFT) of the received signals. The speech signals, denoted $s_j$ for the $j^{\rm th}$ source, undergo acoustic transfer functions, denoted $h_{m,j}$, and are received at microphone $m$ denoted by $x_m$. Let $\mathbf{y} = [y_1, ...., y_{N_m}]^T$, $ \mathbf{x} = [x_1, ...., x_{N_m}]^T , \mathbf{v} = [v_1, ...., v_{N_m}]^T$ and $\mathbf{H}$ be the stacked acoustic transfer function matrix represented by $\mathbf{H} = [\mathbf{h}_1 , \dots, \mathbf{h}_{N_s}]$ where $\mathbf{h}_j$ is a column vector given by $\mathbf{h}_j = [h_{1,j}, \dots, h_{N_m,j}]^T$ for $ j = 1, \dots, N_s$. Let the vector source signals be given by $\mathbf{s} = [s_1, \dots, s_{N_s}]$, therefore the received signal may be expressed as $\mathbf{x} = \mathbf{H}\mathbf{s}$. At the $m^{\rm th}$ microphone the STFT of the received signal is given by, 
\begin{equation}
\begin{split}
y_m [i, k] &= \sum_{j=1}^{N_s}h_{m,j}[k]s_j[i, k] + v_m[i, k] \\
&= x_m[i, k] + v_m[i, k],
\end{split}
\end{equation}
where $i$ and $k$ are the time and frame indices respectively. Here, the multiplicative transfer function assumption is made. In subsequent sections, the dependence on time and frequency is omitted for brevity.

Suppose our microphone signals are preprocessed by a set of $N_p$ random projections, $\mathbf{\Psi_p} \in \mathbb{C}^{N_d \times N_m}$ such that we now observe the action of $\mathbf{\Psi_p}$ on the microphone signals $\mathbf{y}$, where $N_d$ is the number of output dimensions for $p = 1, \dots, N_p$. The projected signals $\mathbf{{y_p}}$ are given by
\begin{equation}
\mathbf{{y_p}} = \mathbf{\Psi_p}\mathbf{y} = \mathbf{\Psi_p}\mathbf{x} + \mathbf{\Psi_p}\mathbf{v}  = \mathbf{{x_p}} + \mathbf{v_p},
\end{equation}
where $\mathbf{{y_p}} = [{y_{p,1}}, ...., {y_{p,N_d}}]^T$, $\mathbf{{x_p}} = [{x_{p,1}}, ...., {x_{p,N_d}}]^T$, and $\mathbf{{v_p}} = [{v_{p,1}}, ...., {v_{p,N_d}}]^T$. Setting any $\mathbf{\Psi_p} = \mathbf{I}_{{M}}$, the $M \times M$ identity matrix, recovers the sensorspace signals. In the case of distributed arrays such projection requires all devices to share signals with each other synchronously. The output signal for each projection $p$ given by ${z_p}$ is obtained by filtering and summing the received projected signals by the filter weights $\mathbf{{w_p}} = [{w_{p,1}}, \dots, {w_{p,N_d}}]^T$, $ \mathbf{{w_p}} \in \mathbb{C}^{N_d}$
\begin{equation}
{z_p} = \mathbf{{w_p}}^H\mathbf{{y_p}}.
\end{equation}
The projected correlation matrix, the speech correlation, and the noise correlation matrix are given by
\begin{equation}
\Phi_\mathbf{{y_p}} = \mathbb{E}\{\mathbf{{y_p}}\mathbf{{y_p}}^H\}, 
\Phi_\mathbf{{x_p}} = \mathbb{E}\{\mathbf{{x_p}}\mathbf{{x_p}}^H\},
\Phi_\mathbf{{v_p}} = \mathbb{E}\{\mathbf{{v_p}}\mathbf{{v_p}}^H\}.
\end{equation}
\noindent
Here, all signals are assumed zero mean. 

\section{Multi-stage Mixture of Compressed Beamformers}
\subsection{Minimum Variance Distortionless Response Beamformer}

The MVDR beamformer is a matched filter with unity gain constraint in the assumed steering direction $\mathbf{g_j}$ to the $j^{th}$ desired signal. The objective for minimization in the MVDR beamformer can be expressed in the projected space as, 
\begin{equation}
J(\mathbf{{w_{j}}}) = \mathbb{E}\{\|\mathbf{{w_{j}}}^H\mathbf{{y}}\| ^2\} \text{\hspace{1mm} such that } \mathbf{{w_{j}}}^H\mathbf{g_j} = 1
\end{equation}
Substituting $\mathbf{{y_p}}=\mathbf{\Psi_p x + \Psi_p v}$ results in the following expression for the 
MVDR beamformer weights, 
$\mathbf{{w_{p,j}}}_{,opt}= \arg\min_\mathbf{{w_{p,j}}} J(\mathbf{{w_{p,j}}})$,
\begin{equation}\label{Beamspace MWF}
\mathbf{{w_{p,j}}}_{,opt}=\frac{(\mathbf{\Psi_p}(\Phi_\mathbf{x} + \Phi_\mathbf{v})\mathbf{\Psi_p}^H)^{-1}\mathbf{\Psi_pg_j}}{\mathbf{g_j^H\Psi_p^H}(\mathbf{\Psi_p}(\Phi_\mathbf{x} + \Phi_\mathbf{v})\mathbf{\Psi_p}^H)^{-1}\mathbf{\Psi_pg_j}}
\end{equation}
\noindent

\begin{figure}
	\centerline{\includegraphics[width=8cm]{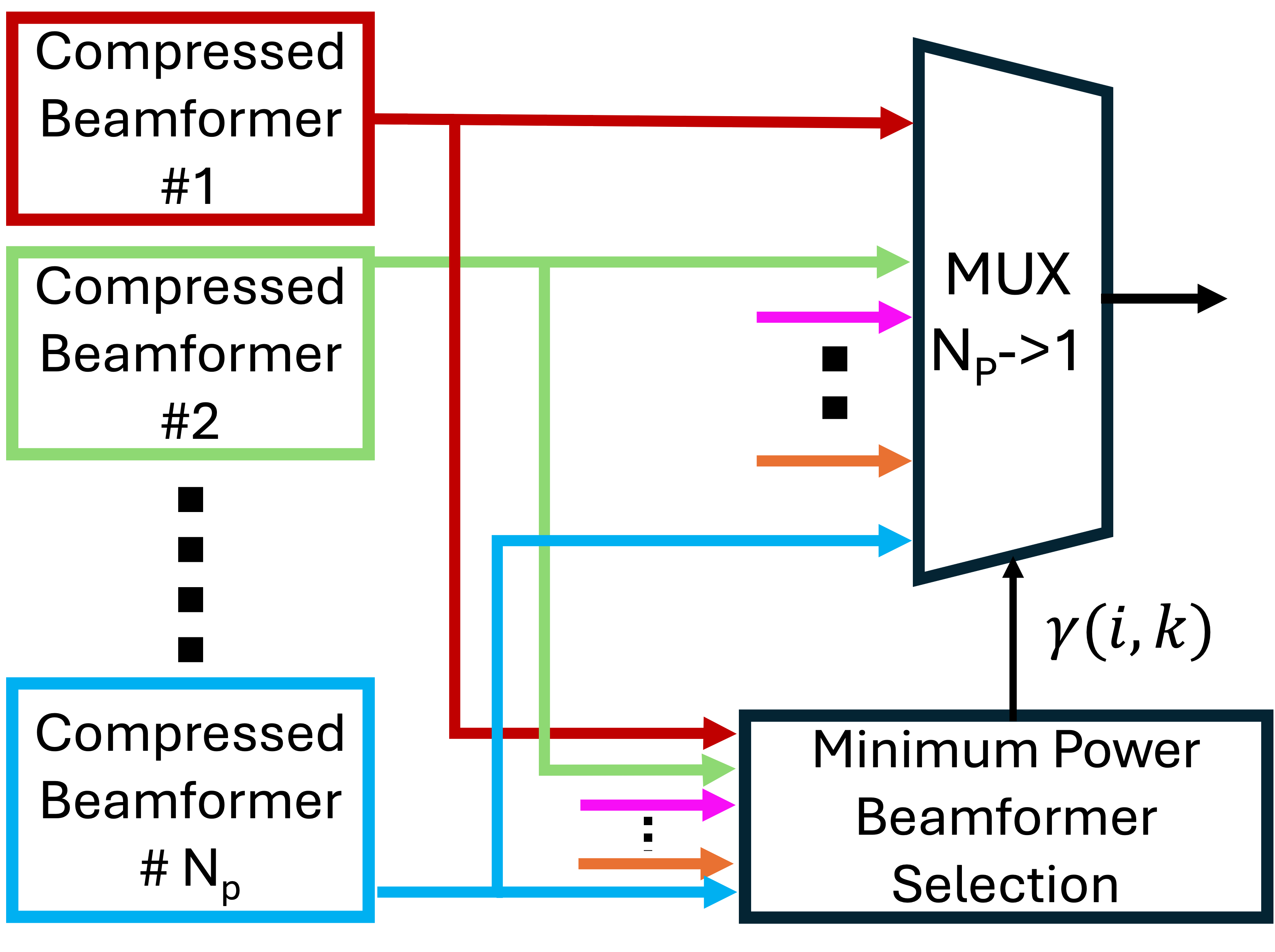}}
	\caption{The time-frequency (TF) sparsity mixture model for the proposed method. To estimate speech signals we exploit TF sparsity and pick the beamformer with the lowest output power. Depicted is the mixture model for a single frequency index $k$ at time step $i$.}
	\label{fig:speech_mixture_model}
\end{figure}

\subsection{Mixture Model}
The universal dominant mode rejection beamformer (UDMR) \cite{udmr} is a performance-weighted blended beamformer that derives its beamforming weights as a blend of dominant mode rejection (DMR) beamformers with different dominant subspace sizes. A key insight in the UDMR beamformer is that the performance blend must rely on the output power of the individual beamformers to use as a proxy for the mean-squared error. A similar observation was made by Corey \cite{ryan_nonstationary} where a time-frequency (TF) sparsity mixture model picks between different MVDR beamformers with a minimum power selection rule. The insight in the UDMR beamformer was that the power arriving from the desired signal direction forms a lower bound on the output power, therefore different beamformers compete to reduce interferer and noise power. Note that this is in line with Capon's criterion for minimum variance beamforming. The TF sparsity mixture model is based on the insight that the MVDR beamformer yields unbiased estimates in a zero-mean noise setting, therefore the mixture model should pick the beamformer with the lowest variance. However, the true variance cannot be measured because we do not have access to the source signal, and must pick the beamformer with the lowest output power. An important distinction exists between the UDMR beamformer and the TF sparsity mixture model. In the UDMR, the accumulated output power of each beamformer is used in a softmax function to derive a blended beamformer, while the TF sparsity chooses the one with minimum power. This difference exists because of the type of signal that is estimated. The UDMR considered white noise with a fixed variance while the TF sparsity model considered speech, which is non-stationary and therefore the mixture does not benefit from the accumulated output power. Our method is equally applicable to both signal estimation scenarios. In this paper we consider speech sources, therefore the decision rule $\gamma[i, k]$ at time index $i$ and frequency index $k$, may be expressed as,
\begin{equation}
    \gamma[i, k] = \arg\!\min_{p}  |z_p[i, k]|^2.
\end{equation}

\noindent
Figure \ref{fig:speech_mixture_model} depicts the mixture model for speech and similarly non-stationary signals.

\subsection{Compressed Beamforming: Upper and Lower Bounds}
Each MVDR beamformer is constrained to have unity gain in the assumed steering direction. This implies that desired signal energy forms a lower bound on the output power of the beamformers. In that sense, each of the beamformers compete to reduce background noise and interferer power. In this section we derive upper and lower bounds for the regret of a single compressed MVDR (CMVDR) beamformer in terms of output power when compared to the sensorspace MVDR beamformer. Therefore, we may express the regret $R$ as,

\begin{equation}
\begin{split}
    R &= p_{CMVDR} - p_{MVDR} \\
    &= (\mathbf{g_j^H\Psi_p^H}(\Phi_\mathbf{y_p})^{-1}\mathbf{\Psi_pg_j})^{-1} - (\mathbf{g_j^H}( \Phi_\mathbf{y})^{-1}\mathbf{g_j})^{-1}
\end{split}
\end{equation}
\noindent
The output power of the full complexity beamformer is easily bounded if we consider the spectral norm of the received correlation matrix. Denoting $\lambda_{max}$ and $\lambda_{min}$ as the maximum and minimum values of $\Phi_{y}$, we have

\begin{equation}
    \frac{\lambda_{min}}{\| \mathbf{g_j} \|^2} \leq p_{MVDR} \leq \frac{\lambda_{max}}{\| \mathbf{g_j} \|^2}
\end{equation}
\noindent
Recall, that if a matrix $\mathbf{\Psi}$ satisfies the restricted isometry property (RIP) of order K, then there exists a $\delta_k > 0$ such that for any K-sparse signal s,

\begin{equation} \label{RIP}
    (1-\delta)\|s\|^2 \leq \|\mathbf{\Psi} s\|^2 \leq (1+\delta)\|s\|^2
\end{equation}
\noindent
Equivalently, all eigenvalues of $\mathbf{\Psi}^H\mathbf{\Psi}$ may be bounded as
\begin{equation}
    (1-\delta) \leq \lambda_i (\mathbf{\Psi}^H\mathbf{\Psi}) \leq (1+\delta)
\end{equation}
\noindent
Note that, for any matrix $\mathbf{A}$, $\mathbf{A^T A}$ and $\mathbf{AA^T}$ share the same eigenvalues. Stated explicity, we have 
\begin{equation}
    (1-\delta) \leq \lambda_i (\mathbf{\Psi}\mathbf{\Psi}^H) \leq (1+\delta)
\end{equation}

In order to bound the eigenvalues of the projected correlation matrix, we make use of the multiplicative version of the Lidksii-Mirsky-Wielandt theorem. 

\begin{theorem}[Lidskii-Mirsky-Wielandt]
Let $\mathbf{A}$ be an $n \times n$ Hermitian matrix and let $\Tilde{\mathbf{A}} = \mathbf{S}^* \mathbf{A}\mathbf{S}$. Then for any indices $1 < i_1 < i_2 < \ldots < i_k \leq n$ such that $\lambda_{i_j} \neq 0$ for $ j= 1, \ldots, k$, we have
\begin{equation}
    \Pi_{j=1}^k \lambda_{n+1 -j}(\mathbf{S}^* \mathbf{S}) \leq \Pi_{j=1}^k \frac{\lambda_{i_j}(\Tilde{\mathbf{A}})}{\lambda_{i_j}(\mathbf{A})} \leq \Pi_{j=1}^k \lambda_{j}(\mathbf{S}^* \mathbf{S})
\end{equation}
Here we assume that the eigenvalues are ordered from largest to smallest such that $\lambda_1 \geq \lambda_2 \geq \ldots \lambda_n.$. The non-negativity of the middle terms follows from $\tilde{A}$ and $A$ having the same number of positive (respectively, negative) eigenvalues.
\end{theorem}

Using the previous theorem, we may bound the eigenvalues of $(\mathbf{\Psi_p}(\Phi_\mathbf{y})\mathbf{\Psi_p}^H)^{-1}$ as follows, 
\begin{equation}
     \frac{1}{(1+\delta)\lambda_{max}} \leq \lambda_i ((\mathbf{\Psi_p}(\Phi_\mathbf{y})\mathbf{\Psi_p}^H)^{-1}) \leq \frac{1}{(1-\delta)\lambda_{min}}
\end{equation}
\noindent
Again, using spectral theory and equation \ref{RIP}, we can bound the output power of the compressed MVDR beamformer, in the following manner,
\begin{equation}
     \frac{(1-\delta)\lambda_{min}}{(1+\delta)\|\mathbf{g_j}\|^2} \leq p_{CMVDR} \leq \frac{(1+\delta)\lambda_{max}}{(1-\delta)\|\mathbf{g_j}\|^2}
\end{equation}
Putting it together, we can bound the \textit{regret} of the compresssed beamformer
\begin{equation}
    \frac{(1-\delta)\lambda_{min}}{(1+\delta)\|\mathbf{g_j}\|^2} - \frac{\lambda_{max}}{\| \mathbf{g_j} \|^2} \leq R \leq \frac{(1+\delta)\lambda_{max}}{(1-\delta)\|\mathbf{g_j}\|^2} - \frac{\lambda_{min}}{\| \mathbf{g_j} \|^2}
\end{equation}

In particular, this implies that any one compressed MVDR may outperform the sensorspace MVDR beamformer. This concludes our proof.
\section{Experiments and Discussion}

A simulated room of size 5 m x 4 m x 4 m is created in Pyroomacoustics \cite{pra}. The maximum order for reflections is set to 10 to ensure that the impulse response from source to receiver is sufficiently dense. Five sources are located at (1m, 1m, 1.8m), (2m, 2m, 1.8m), (3m, 3m, 1.8m), (3m, 1m, 1.8m), and (1m, 3m, 1.8m) respectively. The sampling rate is set to 16kHz. The speech signals are derived from the CMU ARCTIC speech database \cite{kominek2004cmu}. Three ceiling mounted circular microphone arrays of radius 0.18m, with 10 elements each, are positioned at (1.5m, 1.4m, 4m), (2.5m, 3m, 4m), and (3.5m, 1m, 4m). The array is circular and the array radius was roughly chosen to be matched at 1.5 kHz. For matched spacing, the inter-element spacing $d = \frac{c}{2\times1500}$, where $c$ denotes the speed of sound, assumed to be 343 m/s. The array radius $R$, is then given by,

\begin{equation}
    R = \frac{d}{2\sin(\pi/N_m)}
\end{equation}

\begin{figure}
    \centerline{\includegraphics[width=9cm]{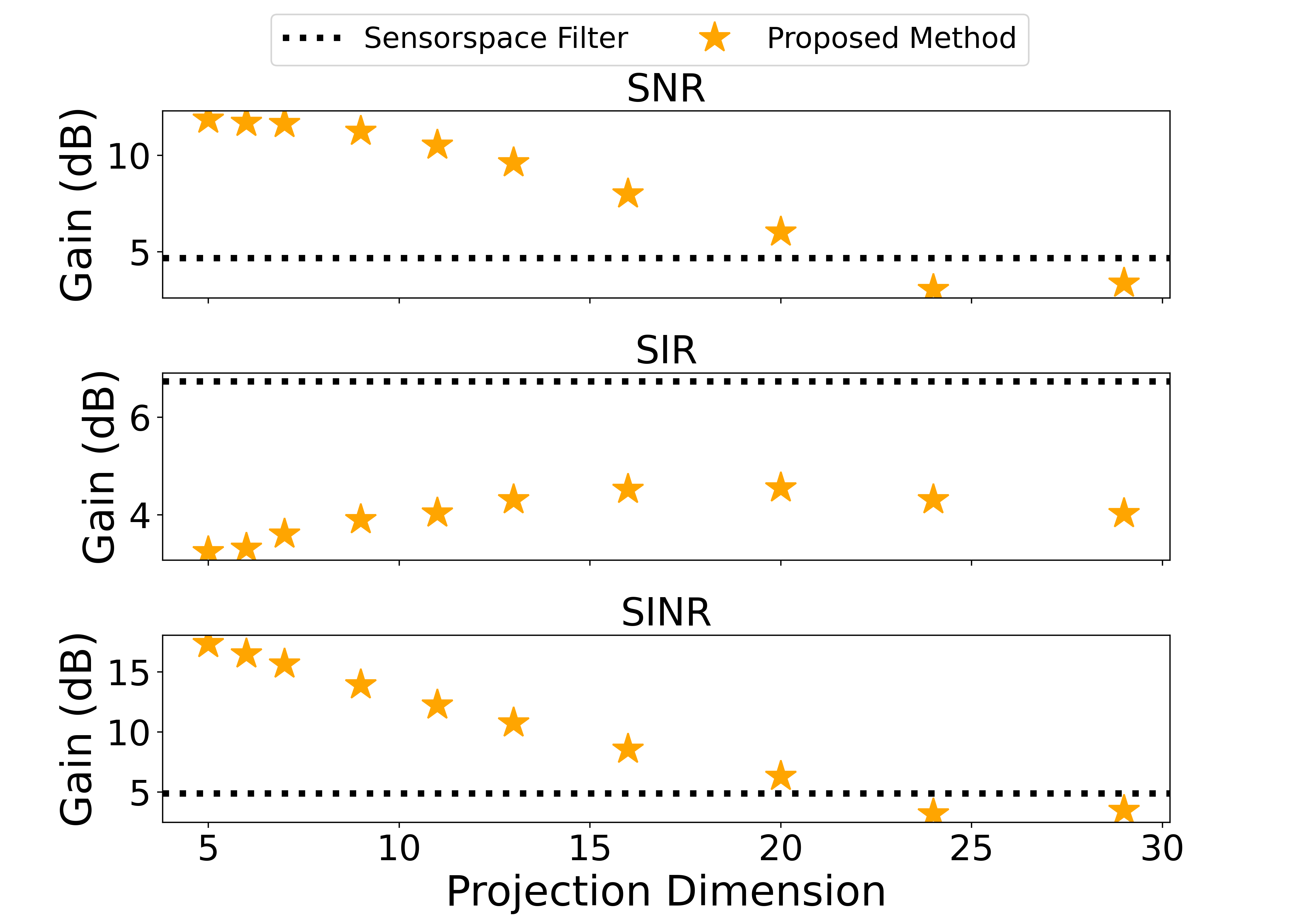}}
 \caption{The SNR, SIR, SINR gain in dB for the sensorspace MVDR and the performance weighted-blended compressed MVDR beamformer for talker 1.}
	\label{fig:t1_metrics}
\end{figure} 
To test the performance of the proposed method against the sensorspace MVDR beamformer the number of dimensions in each projection $N_d$ is varied from 5 to 30 using a log spacing sampled 10 times to generate the results. The number of projections $N_p$ is limited so that $N_p N_d^3 \leq N_m^3$. As the number of dimensions per projection increases, the cost of projection can become significant, and this is observed in the results as the number of projections subsequently reduces to 1. At such points, projection is sub-optimal when constrained to have lower computational complexity. At smaller $N_d$, the method has the capacity to support numerous projections and take advantage of the compressed beamformers. As such the filters are trained and tested on different recordings. The channel is estimated in sensorspace and each compressed beamformer uses the same random projection to generate its own channel estimate. Figure \ref{fig:t1_metrics} shows the signal-to-noise ratio (SNR), signal-to-interferer ratio (SIR), and signal-to-interferer-and-noise ratio (SINR) gains in dB for talker 1, when using Gaussian random projections. The figure demonstrates that the proposed method is able to more reliably provide better SNR and SINR gain when using a large range for $N_d$. 

Our method had a similar trend for each of the remaining 4 talkers. We can see that the proposed decomposition quickly approaches SINR gain for the full-complexity MVDR beamformer and quickly surpasses it. Notably this is achieved with similar computational complexity. Since the training period is used to design a linear and time-invariant beamformer, we can understand the performance improvement of our algorithm in the context of a rich history of work in universal algorithms \cite{singer, merhav, 4244698, 4359533, udmr, 10051929, 9977261, 10636392,zhuang2024active}. In that work, the ability to switch or blend between different candidate algorithms in a set, allows us to achieve a performance that is better than any of the constituent algorithms. Similarly, switching between multiple compressed beamformers allows us to achieve a better performance than any of the individual beamformers. The time-adaptive choice also enables us to outperform the sensorspace MVDR. We notice that our current approach presents a model order problem in itself similar to \cite{singer,zhuang2024active}. Future work will focus on efficient recursive decomposition of the compressed beamformer to adaptively pick the projection dimension too.

\section{Conclusion}
In this paper, we aimed to extend the compressed signal processing framework to beamformers. We elected to use random projections to embed the sensor signals in a low dimensional space. To account for the distortion induced by one such projection, we employ multiple projections in parallel and combine their outputs using a TF sparsity mixture model. In our experiments, the performance of the mixture compressed MWF is compared to the sensorspace MWF. When the relative computational complexity, that is $N_p = (\frac{N_m}{N_d})^3$, is held constant and as the embedding dimension gets smaller, the number of projections the method is allowed to use increases significantly. This introduces a tradeoff between the number of projections and the distortion in each embedding. With fewer projections, we require that each projection have a lower distortion by itself. While each embedding with small dimensions will have a larger distortion, we can compensate for the loss with a large number of projections. The mixture algorithm is essential in finding the balancing point of this tradeoff for each embedding dimension. Additionally, our method can be implemented efficiently in parallel, meaning that a system designer could bring the latency down to that of a single compressed beamformer with a small overhead for the mixture algorithm. With today's fast parallel processors, our method provides a way of achieving improved performance and latency. 
\section*{Acknowledgement}
Funded in part by US Navy, Office of Naval Research under award N00014-23-1-2133

\bibliography{ieee-bibliography}
\vspace{12pt}
\color{red}

\end{document}